\documentclass[prb,groupedaddress,twocolumn,preprintnumbers]{revtex4-1}
\usepackage{amsfonts, amsmath, amssymb, bm, bbm, slashed}
\usepackage{caption, subcaption}
\usepackage{graphicx}
\DeclareMathOperator{\tr}{Tr}
\DeclareMathOperator{\im}{Im}
\DeclareMathOperator{\re}{Re}
\renewcommand{\vec}[1]{{\mathbf{#1}}}
\newcommand{\beq}{\begin{eqnarray}}
\begin{document}

\title{Minding the Gap in Holographic Models of Interacting Fermions}
\author{Garrett Vanacore and Philip W. Phillips}
\affiliation{Department of Physics, University of Illinois at Urbana-Champaign,
Urbana, IL 61801} 
\date{\today}

\begin{abstract}
We study the holographic dual of fermions interacting in a Schwarzschild-AdS$_{d+1}$  background via a dipole (Pauli) coupling sourced by a probe gauge field.  We find quite generally
that a gap forms in the dual operator spectrum as the Pauli coupling is strengthened. Previous investigations have observed this behavior in analogous constructions with Reissner-Nordstr\"om-AdS (RN-AdS$_4$) backgrounds, but the emergence of log-oscillatory behavior in those models' spectra prevented identification of the underlying gapping mechanism. Our model obviates this issue through its modified geometry and traces the gapping mechanism back to the bulk dynamics.  We show in general that there is a duality between zeros for large positive values of the coupling and poles in the spectrum for equivalent couplings but with opposite sign as seen recently in the RN-AdS$_4$ background\cite{alsup}.  The duality arises from the two possible quantizations for computing the retarded propagator.   Coupled with the earlier string results\cite{gauntlett,gubser2} that Fermi surfaces are generally absent from the spectral function, our finding that the Pauli term engineers the gap suggests that the model examined here offers a way of studying non-perturbative physics in fermionic matter at finite density typified by Mott insulating systems.
\end{abstract}

\maketitle

%%%%%%%%%%%%%%%%%%%%%%
Dynamical generation of a mass gap is a feature that typifies many strongly correlated systems.  In quantum-chromodynamics in $1+1$ dimensions \cite{thooft},  
there are no physical quark states at low energy because the pole in their propagator
moves off to infinity.  In the Mott problem, the real and imaginary parts of the electron propagator vanish identically at zero energy\cite{dzy} although the charge  density
is non-zero.  We are concerned 
in this paper primarily with models that generate zeros of the single-particle propagator as exemplified by the Mott problem.  Because the charge gap in Mott systems obtains without any change
in the size of the Brillouin zone, standard mean-field approaches are not applicable and strongly coupled techniques must be adopted.  In this regard, dynamical mean-field theory  \cite{dinfty} has been highly successful even yielding a surface of zeros  \cite{kotstan,imada} for the Hubbard model in $d=2+1$ dimensions.   Since this method is approximate, what would be preferable is an exact method in the difficult non-perturbative parameter space where zeros of the single-particle propagator reside.   

In this context, the Gauge-gravity duality  \cite{maldacena}, which offers a precise dual mapping between a strongly coupled $d$-dimensional quantum theory and a weakly interacting theory of gravity  in $d+1$ dimensions, offers a potential way forwards.   In actuality, it is difficult to construct the gravity equivalent of specific quantum theories, for example the Hubbard model.  Nonetheless, progress can still be made because the details of the bulk theory of gravity can be tuned to generate desired characteristics of the boundary correlators on the grounds  \cite{witten,gubser}  that fields in the bulk act as sources for operators at the boundary.   Within this framework, known as bottom-up constructions, Edalati, et al.  \cite{edalati1,edalati2} showed that the boundary fermionic propagator for a bulk theory involving the Pauli coupling in a Reissner-Nordstr\"om anti de Sitter spacetime produces a spectrum which has vanishing spectral weight at a range of energies around $\omega=0$ without the breaking of any symmetry.  Consequently, this gap-like feature was argued to be reminiscent of the Mott gap.  A  curious feature of this result is that the gap appeared to be due to the emergence of a lower AdS$_2$ symmetry associated with the near horizon region of the background.  In fact, the gap-like feature obtains only when the Pauli term exceeds a critical value such that the Fermi momentum is pushed into the parameter space where the Green function becomes discrete scale invariant as a function of frequency, also known as  log-oscillatory behavior.  In addition, Edalati, et al.  \cite{edalati1,edalati2} argued that the gap-like feature arises from a vanishing of the quasiparticle weight.  This result was corroborated  \cite{gauntlett,gubser} by an exact top-down construction which showed that the Pauli term is a generic feature of such theories and the resultant spectral function exhibits a dip at the chemical potential and no Fermi surface.   Also, extensive bottom-up constructions  \cite{dipole1,dipole2,dipole3,dipole4,dipole5,dipole6} have found a similar gapping tendency of the Pauli term.   However, recently a rather intriguing result  \cite{alsup} has been obtained which sheds new light on the origin of the spectral weight suppression that results from the Pauli term in an RN-AdS$_4$ background.   Alsup, et al.  \cite{alsup} showed that the flow equations admit the duality
	\begin{align*}
		 {\rm Det G_R}(\omega=0,k;p)=\frac{1}{{\rm Det G_R}(\omega=0,-k;-p)},
	\end{align*}
where $p$ is the magnitude of the Pauli term.
Because the propagator for extreme negative values of $p$ exhibits only poles, the corresponding propagator for large positive values of $p$ contains only zeros.  Consequently, the vanishing of the spectral weight is due to zeros and the bulk Pauli coupling in RN-AdS$_4$ mimics Mott physics.  

In this paper, we re-investigate the role of a bulk Pauli coupling in gravity constructions with an eye to addressing two problems, namely the role of the log-oscillatory region and the precise origin of the zeros-pole duality.   To address these questions, we use a Schwarzschild background because it does not possess a near-horizon region that gives rise to log-oscillatory physics.   To achieve a finite density description, we include a gauge field that sources a chemical potential at the boundary and calculate the resultant propagators.  We find that the spectrum is still gapped and hence, the origin of the gap is independent of discrete scale invariance.  We also find a zeros-pole duality and show that it is a general consequence of the standard and alternative quantizations  \cite{witten,gubser} for computing the retarded propagator in holography.  

\section{Bulk Model}

Our model parallels the earlier work of Edalati, et al.  \cite{edalati1}, but differs in our choice of Schwarzschild-AdS$_{d+1}$ (SS-AdS) as the background geometry. We consider spin-$1/2$ fermions coupled to a gauge field through a dipole interaction $F_{ab} \bar{\psi} \Gamma^{ab} \psi$. The corresponding bulk Lagrangian is
	\begin{align*}
		\sqrt{-g} i \bar{\psi} \left( \slashed{D} - m - i p \slashed{F} \right) \psi,
	\end{align*}
where
	\begin{align}
		\bar{\psi} &= \psi \Gamma^{\underline{t}}, \nonumber \\
		\slashed{D} &= e_c^M \Gamma^c \left( \partial_M + \frac{1}{4} \omega_M^{ab} \Gamma_{ab} - i q A_M \right), \\
		\slashed{F} &= \frac{1}{2} \Gamma^{ab} e_a^M e_b^N F_{MN}, \nonumber
	\end{align}
with $e_a^M$ and $\omega_M^{ab}$ being the inverse frame field and spin connection (respectively). Our index notation uses capital letters for bulk coordinates $M,N \cdots = \{ t,x^i, r \}$, lower case letters for tangent space coordinates $a, b, \cdots = \{ \underline{t}, \underline{x}^i, \underline{r} \}$, and Greek letters $\mu, \nu, \cdots$ for boundary coordinates. Dirac matrices $\Gamma^{\underline{t}}, \Gamma^{\underline{1}}, \cdots, \Gamma^{\underline{r}}$ are chosen such that the Clifford algebra $\{ \Gamma^a, \Gamma^b \} = 2 \eta^{ab}$ is satisfied. Additionally $\Gamma_{ab} = \tfrac{1}{2} [\Gamma_a, \Gamma_b]$. In our calculations, we scale the Pauli coupling as $p \rightarrow pL/(d-2)$ for convenience.

We parametrize the geometry of the SS-AdS background in the Poincar\'e patch, writing
	\begin{align} \label{eq:metric}
		ds^2 = \frac{r^2}{L^2} \left[ - f(r) dt^2 + d \vec{x}^2 \right] + \frac{L^2}{r^2} \frac{dr^2}{f(r)},
	\end{align}
where the metric function is given by
	\begin{align} \label{eq:ssmetfn}
		f(r) = 1 - M \left( \frac{r_0}{r} \right)^d.
	\end{align}
The mass and horizon of the black hole are respectively denoted by $M$ and $r_0$, with $r_0$ being determined by the largest positive root of $f(r_0) = 0$. The temperature of the black hole, which determines the temperature of the dual theory, is given by
	\begin{align}
		T =  \frac{r_0}{4 \pi L^2} \left( 2 + (d-2)M \right).
	\end{align}
Since our focus is matter at finite density, a gauge field is essential to our model but is not naturally supported by SS-AdS. To remedy this, we introduce a probe gauge field parametrized like that appearing in RN-AdS,
	\begin{align}
		A &= \mu \left[1 - \left( \frac{r_0}{r} \right)^{d-2} \right] dt, \label{eq:gaugefield} \\
		\mu &= \left( \frac{d-1}{2d-4} \right)^{1/2} \frac{Q r_0}{L^2},
	\end{align}
and continue to interpret $\mu$ as the chemical potential of the dual theory.  Note that $Q$, which originally represents the charge of the RN black hole, is treated simply as a tunable parameter throughout our analysis.

It may seem arbitrary at first to formulate this model in SS-AdS with a gauge probe rather than RN-AdS, but doing so simplifies analysis of the dual theory by removing certain complicating features of RN-AdS$_4$.It is well known that the near-horizon geometry of RN-AdS$_4$is quite special, becoming AdS$_2 \times \mathbb{R}^{d-1}$ in the extremal case or a black hole in AdS$_2$ in the finite temperature case. As a consequence, any field introduced in the bulk sources not only a UV operator in the asymptotic boundary, but also a tower of operators in the near-horizon region  \cite{faulkner}. The emergence of this IR CFT is reflected directly in the small frequency behavior of the UV operators, notably inducing log-oscillatory behavior in boundary two-point functions. In contrast, SS-AdS removes the black hole charge and, by extension, the length scale necessary for the emergent AdS$_2$ geometry. Thus the complication of log-oscillatory behavior should not occur for operators in SS-AdS, and ideally we can study other phenomenology (such as gapping) down to zero frequency.  The complete details of how the correlators are determined are contained in the Appendix.

The equations of motion of the bulk fermions are most easily analyzed in momentum space, so we Fourier transform the fields as $\psi(r, x^{\mu}) \sim e^{i k \cdot x} \psi(r,k^{\mu})$, where $k^{\mu} = (\omega, \vec{k})$. The Fourier-transformed Dirac operator $\slashed{D}$ is
	\begin{align}
		\slashed{D} &= \frac{r}{L} \sqrt{f(r)} \Gamma^{\underline{r}} \left[ \partial_r + \frac{f'(r)}{4f(r)} + \frac{d}{2r} \right] \nonumber \\
		&\quad -i \frac{L}{r \sqrt{f(r)}} \Gamma^{\underline{t}} \left[ \omega + q A_t (r) \right] + i \frac{L}{r} \vec{k} \cdot \vec{\Gamma},
	\end{align}
and the field strength is
	\begin{align}
		\slashed{F} = (d-2) \frac{\mu}{r_0} \left( \frac{r_0}{r} \right)^{d-1} \Gamma^{\underline{r}\underline{t}}.
	\end{align}
The equations of motion may be decoupled by introducing projectors $\Gamma_{\pm} = \tfrac{1}{2} (1 \pm \Gamma^{\underline{r}} \Gamma^{\underline{t}} \hat{k} \cdot \vec{\Gamma})$ and writing the fields as $\psi_{\pm} (r) = r^{d/2} f(r)^{1/4} \Gamma_{\pm} \psi(r)$.  Due to the rotational symmetry of the spatial boundary coordinates we may set $k_1 = k$ and $k_{i \neq 1} = 0$. An appropriate basis of Dirac matrices is then
	\begin{align}
		\Gamma^{\underline{r}} &=
			\begin{pmatrix}
				- \sigma_3 \otimes \mathbbm{1} & 0 \\
				0 & - \sigma_3 \otimes \mathbbm{1}
			\end{pmatrix}, \nonumber \\
		\Gamma^{\underline{t}} &=
			\begin{pmatrix}
				i \sigma_1 \otimes \mathbbm{1} & 0 \\
				0 & i \sigma_1 \otimes \mathbbm{1}
			\end{pmatrix}, \\
		\Gamma^{\underline{1}} &=
			\begin{pmatrix}
				- \sigma_2 \otimes \mathbbm{1} & 0 \\
				0 &  \sigma_2 \otimes \mathbbm{1}
			\end{pmatrix}, \nonumber
	\end{align}
where the $\sigma_i$ are Pauli matrices and $\mathbbm{1}$ is an identity matrix of dimension $2^{(d-3)/2}$ or $2^{(d-4)/2}$ for $d$ odd or even (respectively). With these choices the equations of motion are
	\begin{align} \label{eq:eom}
		&\frac{r^2}{L^2} \sqrt{f(r)} \partial_r \psi_{\pm} = \frac{i \sigma_2}{\sqrt{f(r)}} \left[ \omega + \mu q \left( 1 - \frac{r_0^{d-2}}{r^{d-2}} \right) \right] \psi_{\pm} \nonumber \\
		&\quad - \sigma_1 \left( \mu p \frac{r_0^{d-2}}{r^{d-2}} \pm k \right) \psi_{\pm} - \sigma_3 \frac{r}{L} m \psi_{\pm}
	\end{align}
which have the asymptotic solutions
	\begin{align} \label{eq:asymp_soln}
		\psi_{\pm} (r, \omega, k) &= a_{\pm} (\omega, k) r^{mL} 
		\begin{pmatrix}
			0 \\
			1
		\end{pmatrix}
		[ 1 + \dots ] \nonumber \\ 
		&\quad + b_{\pm} (\omega, k) r^{-mL} 
		\begin{pmatrix}
			1 \\
			0
		\end{pmatrix}
		[ 1 + \dots ]. 
	\end{align}
When $m \in [0,\frac{1}{2})$ both components in \eqref{eq:asymp_soln} are normalizable, giving one the freedom to choose either $a_{\pm}$ or $b_{\pm}$ as sources for the dual fermion operator in the boundary theory. We will take $m$ to be in this range and work with the conventional quantization where $a_{\pm}$ are the sources. The dual fermion operator then has dimension $\Delta = \frac{d}{2} + mL$.  As we will see, the zeros-pole duality is fundamentally rooted in the choice of the quantization.  Imposing in-falling boundary conditions near the horizon gives a retarded two-point function of the form
	\begin{align} \label{eq:greenfn}
		G_{R}(\omega,k) &=
		\begin{pmatrix}
			G_+(\omega,k) \mathbbm{1} & 0 \\
			0 & G_- (\omega,k)  \mathbbm{1}
		\end{pmatrix}
	\end{align}
where $G_{\pm} (\omega,k) = b_{\pm} (\omega,k)/a_{\pm} (\omega,k)$.

The Dirac equation \eqref{eq:eom} is analytically intractable for general values of $\omega$ and $k$, so we must resort to numerical methods to study the full spectrum of the boundary theory. It is convenient to first cast the equation in dimensionless variables by scaling $r$, $ \omega$, and $k$ as
	\begin{align}
		r \rightarrow r_0 u, \quad \omega \rightarrow \frac{r_0}{L^2} \omega, \quad k \rightarrow \frac{r_0}{L^2} k.
	\end{align}
The rescaled form of \eqref{eq:eom} can then be cast as non-linear flow equations  \cite{faulkner,edalati1}.   Writing the bulk spinors as $\psi_{\pm} = (\beta_{\pm}, \alpha_{\pm})^T$ and defining $\xi_{\pm} = \beta_{\pm} / \alpha_{\pm}$ leads to the flow equation
	\begin{align} \label{eq:flow}
		u^2 \sqrt{f(u)} \partial_u \xi_{\pm} &= - 2 (mL) u \xi_{\pm} + \left[ v_- (u) \mp k \right] \nonumber \\
		&\qquad + \left[ v_+ (u) \pm k \right] \xi_{\pm}^2,
	\end{align}
where
	\begin{align} \label{eq:v}
		v_{\pm} (u) = \frac{1}{\sqrt{f(u)}} \left[ \omega + Qq(1 - u^{2-d}) \right] \pm Q p u^{2-d}.
	\end{align}
With the new function  $\xi_{\pm}$, we find that the in-falling boundary conditions on $\psi_{\pm}$ translate to 
	\begin{align} \label{eq:ssbcs}
		\xi_{\pm} \big|_{u=1} = i \qquad \omega \neq 0.
	\end{align}
The boundary conditions for $\omega = 0$ have a complicated form that we derive in the Appendix, but we note that they are real for the results discussed below. The two-point correlator \eqref{eq:greenfn} takes the form
	\begin{align} \label{eq:xigreenfn}
		G_{R}(\omega,k) &= \lim_{\epsilon \rightarrow 0} \epsilon^{-2mL}
		\begin{pmatrix}
			\xi_+ \mathbbm{1} & 0 \\
			0 & \xi_-  \mathbbm{1}
		\end{pmatrix} \Big|_{u=\frac{1}{\epsilon}},
	\end{align}
where finite terms should be retained as $\epsilon \rightarrow 0$. The spectrum of the boundary theory is given by the fermion spectral function, which is defined up to normalization as
	\begin{align}
		A(\omega,k) = \tr \im G_R (\omega,k).
	\end{align}

\section{Gap Formation and Characterization}

\begin{figure*}
	\centering
	\begin{subfigure}[b]{0.49\linewidth}
		\centering
		\includegraphics[scale=0.6]{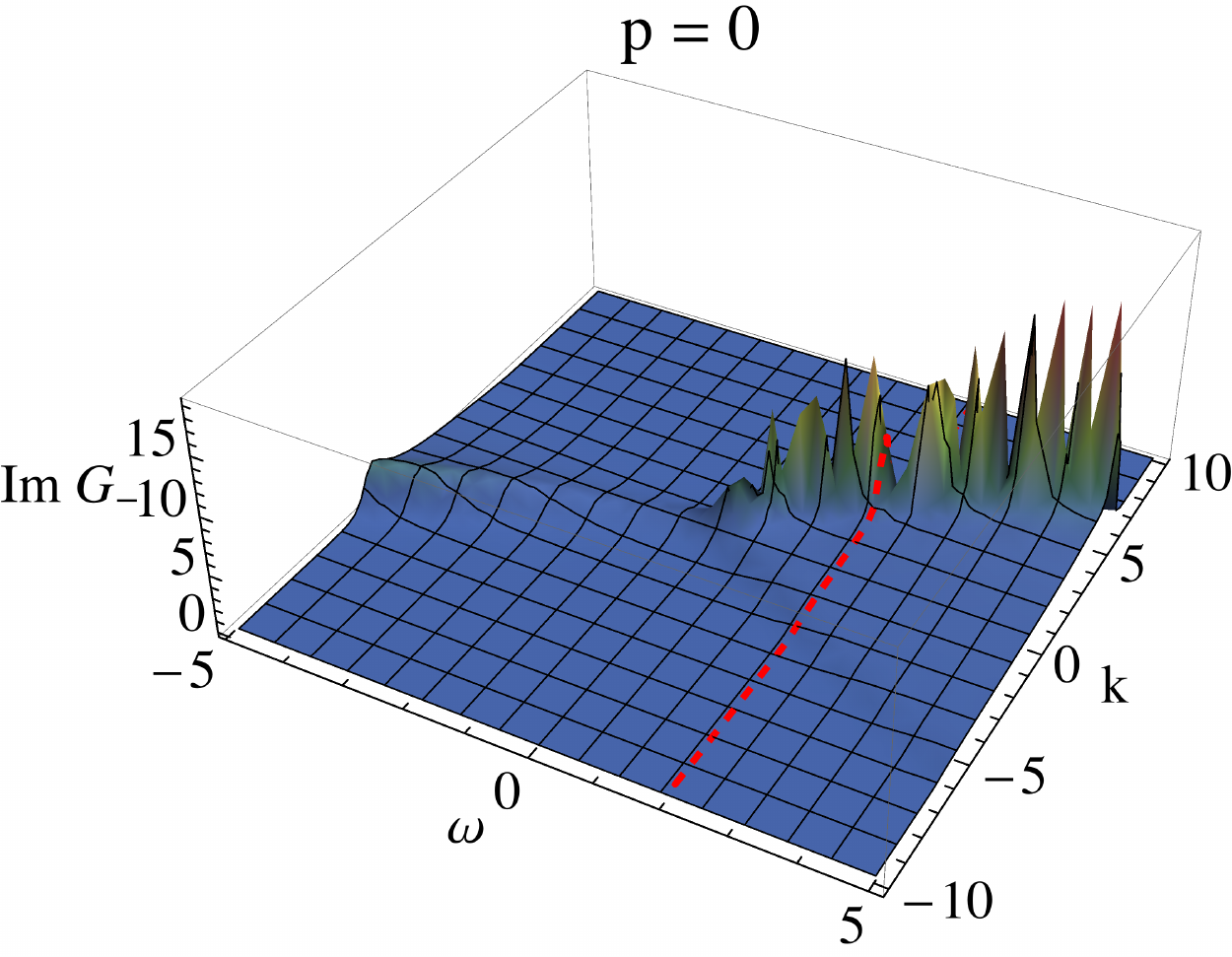}
	\end{subfigure}
	~
	\begin{subfigure}[b]{0.49\linewidth}
		\centering
		\includegraphics[scale=0.6]{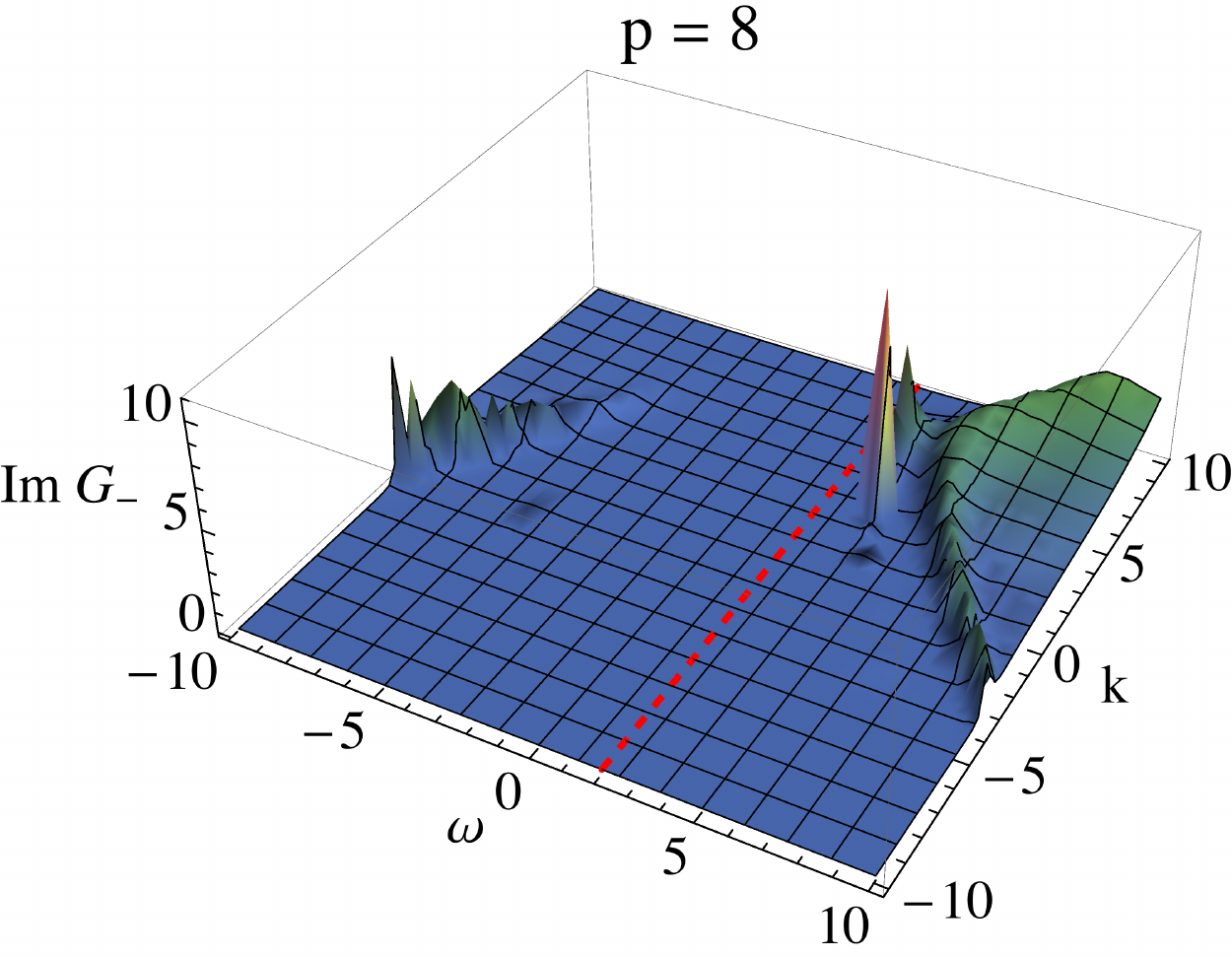}
	\end{subfigure}
	\captionsetup{justification=raggedright,singlelinecheck=false}
		\caption{\footnotesize $\im G_- (\omega,k)$ for $p=0$ (left) and $p=8$ (right). The red lines highlight the $\omega=2$ line discussed in Figure \ref{fig:w2}. As can be seen in the right plot, a gap forms as the coupling is strengthened. Fixing the chemical potential via $Q = \sqrt{3}$ places the minimum at $\omega \approx 0.44$ and the critical coupling at $p \approx 6.82$. The contribution of $G_+ (\omega, k)$ may be deduced by taking $k \rightarrow -k$ in the above plots.}
	\label{fig:gap}
\end{figure*}

We systematically studied the excitation spectrum of the boundary theory through numerical integration of \eqref{eq:flow} with the aforementioned boundary data \eqref{eq:ssbcs}. We fix the boundary dimension to $d=3$ and for simplicity set $q=M=1$ and $m=0$. For the sake of comparison with previous studies of the extremal Reissner-Nordstr\"om case  \cite{edalati1,alsup} we fix the chemical potential via $Q = \sqrt{3}$, but we note that simulations with smaller $Q$ are qualitatively identical and differ only in the $p$ value where novel features emerge.

Figure \ref{fig:gap} displays the contribution of $\im G_- (\omega, k)$ to the boundary spectral function at couplings of $p = 0$ and $p = 8$ .   The $p=0$ result confirms the Fermi surface calculations of Faulkner, et al.  \cite{faulkner}.  However, by the time $p$ has been increased to $p=8$, the spectral weight has dropped precipitously to zero over a wide range of energies.  A gap clearly forms in the spectrum as $p$ is increased. Defining the gap by suppression of spectral weight below $10^{-2}$, the critical coupling for the gap formation is $p \approx 6.82$ and its minimum occurs at $\omega \approx 0.44$.   A similar contribution to the spectrum is provided by $G_+ (\omega, k)$, which may be deduced by taking $k \rightarrow -k$ as follows from the $G_+(\omega, k) = G_- (\omega, -k)$ symmetry of the flow equations \eqref{eq:flow}. We omit this contribution for clarity of presentation.

\begin{figure*}
	\centering
	\begin{subfigure}[b]{0.49\linewidth}
		\centering
		\includegraphics[scale=0.6]{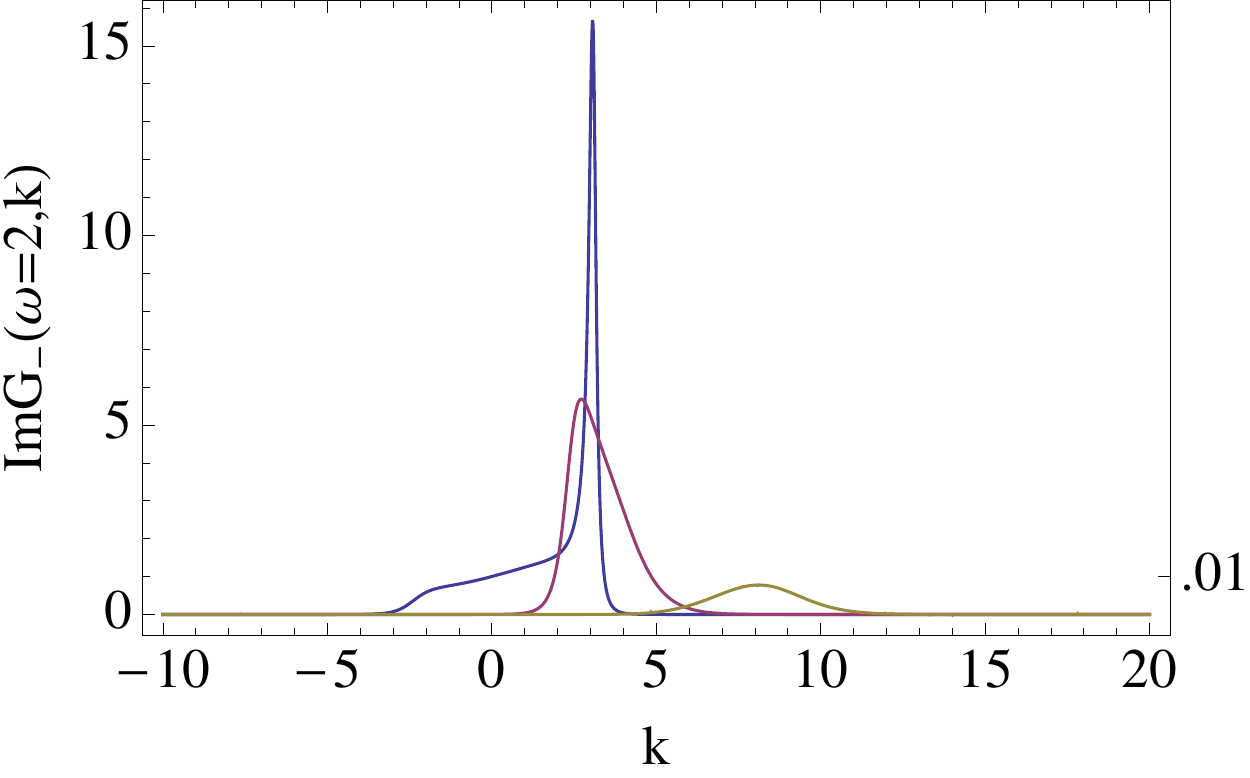}
	\end{subfigure}
	~
	\begin{subfigure}[b]{0.49\linewidth}
		\centering
		\includegraphics[scale=0.6]{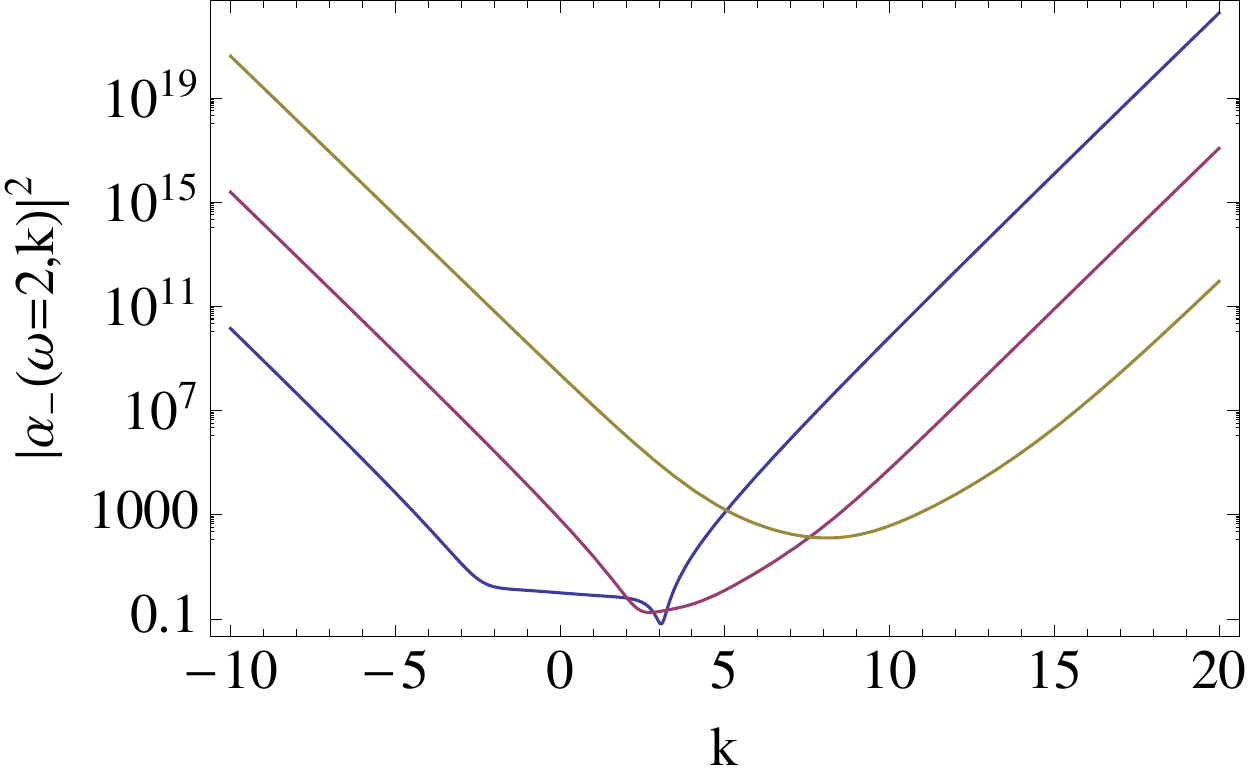}
	\end{subfigure}
	\captionsetup{justification=raggedright,singlelinecheck=false}
		\caption{\footnotesize The $G_-$ contribution to the spectral function (left) and a logarithmic plot of the denominator of $G_-$ (right) at $\omega=2$ for $p=0$ (blue), $p=4$ (red), and $p=8$ (yellow). From the left panel it is clear that strengthening $p$ leads to a strong suppression of spectral weight (note that the $p=8$ plot adheres to the scale on the right and never exceeds 0.01). The behavior of the denominator of $G_-$ (right) correlates exactly with the location of peaks in $\im G_-$ and explains their eventual suppression.}
	\label{fig:w2}
\end{figure*}

The presence of a gap in the Schwarzschild background strongly suggests that the similar findings in the Reissner-Nordstr\"om   \cite{edalati1,alsup} setting are due entirely to the Pauli coupling and not to any discrete scale invariance arising from the log-oscillatory region.  To further investigate this we performed a more detailed study of the bulk spinor components $\alpha_- (\omega,k)$ and $\beta_- (\omega, k)$ along the $\omega=2$ line highlighted in Fig. \ref{fig:gap}.  As can be seen from \eqref{eq:xigreenfn} and the definition of $\xi_-$, these components give more detailed information about the spectral function via
	\begin{align} \label{eq:compimg}
		\im G_- (\omega, k) = \frac{\im \beta_- \re \alpha_- - \re \beta_- \im \alpha_-}{|\alpha_-|^2} \bigg|_{u \rightarrow \infty}
	\end{align}
and an identical expression for $\im G_+ (\omega, k)$. Prior investigations into the Pauli coupling model were unable to examine these details due to the coincidence of the log-oscillatory region with the gap. The spectral function along the line $\omega=2$ as a function of $k$ is shown in Fig. \ref{fig:w2} for $p=0$ (blue), $p=4$ (red) and $p=8$ (yellow).  The spectral weight (first panel in Fig. \ref{fig:w2}) decreases precipitously with increasing $p$, reaching values no larger than .01 regardless of momenta for $p=8$.   In the second panel in Fig. \ref{fig:w2}, we plot the denominator of \eqref{eq:compimg}  for couplings of $p=0$, $p=4$, and $p=8$, with $\omega = 2$.  Focusing on the region between $k\in(-3,3)$, we see that $|\alpha_-|^2$ starts out at a  value of roughly unity for $p=0$ and jumps to a value of $10^{10}$ for $p=8$.  That is, the denominator of the retarded Green function is diverging.  This is the Mott mechanism for gap formation, namely a divergence of the self energy.
\begin{figure*}
	\begin{subfigure}[b]{0.49\linewidth}
		\includegraphics[scale=0.6]{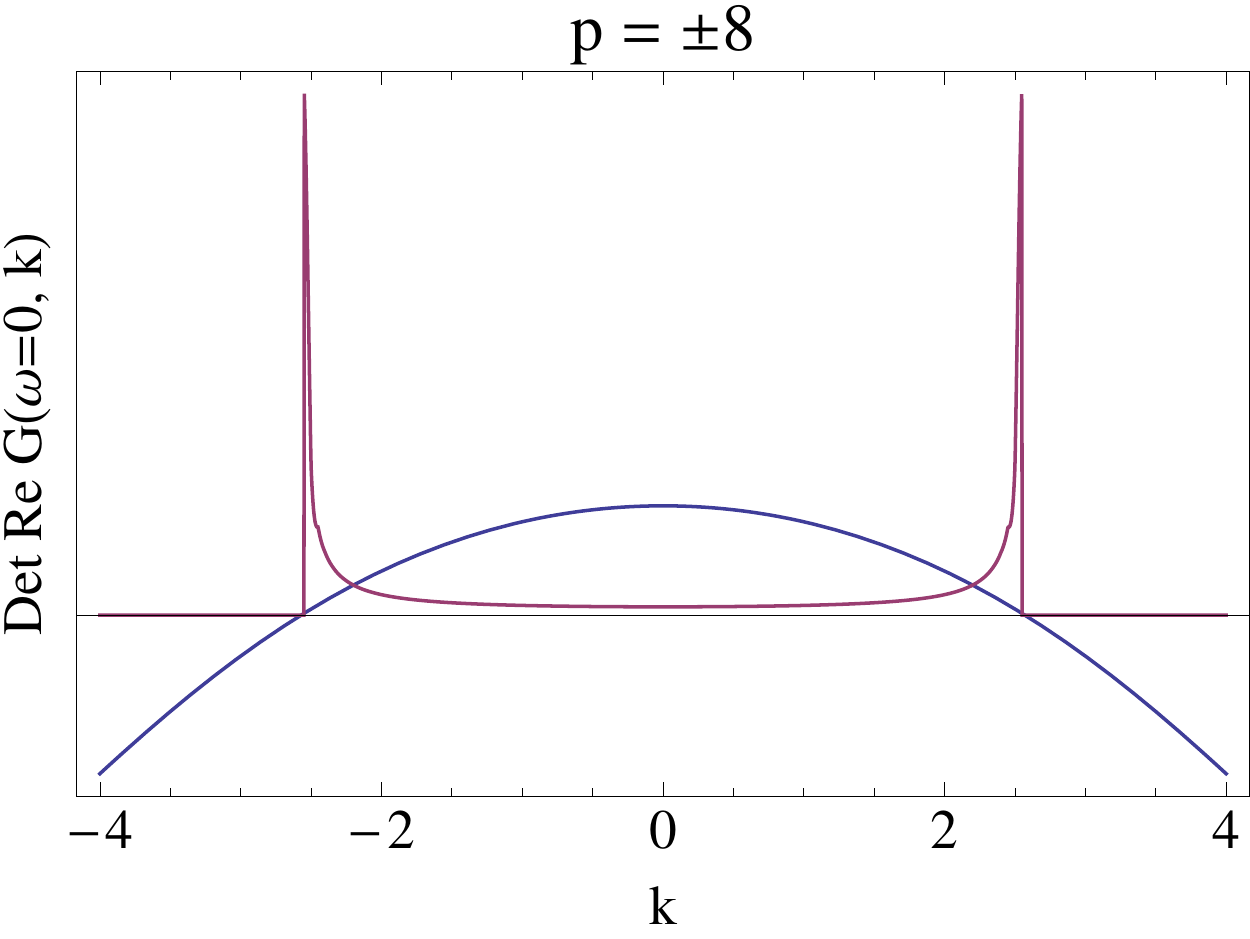}
	\end{subfigure}
	~
	\begin{subfigure}[b]{0.49\linewidth}
		\includegraphics[scale=0.6]{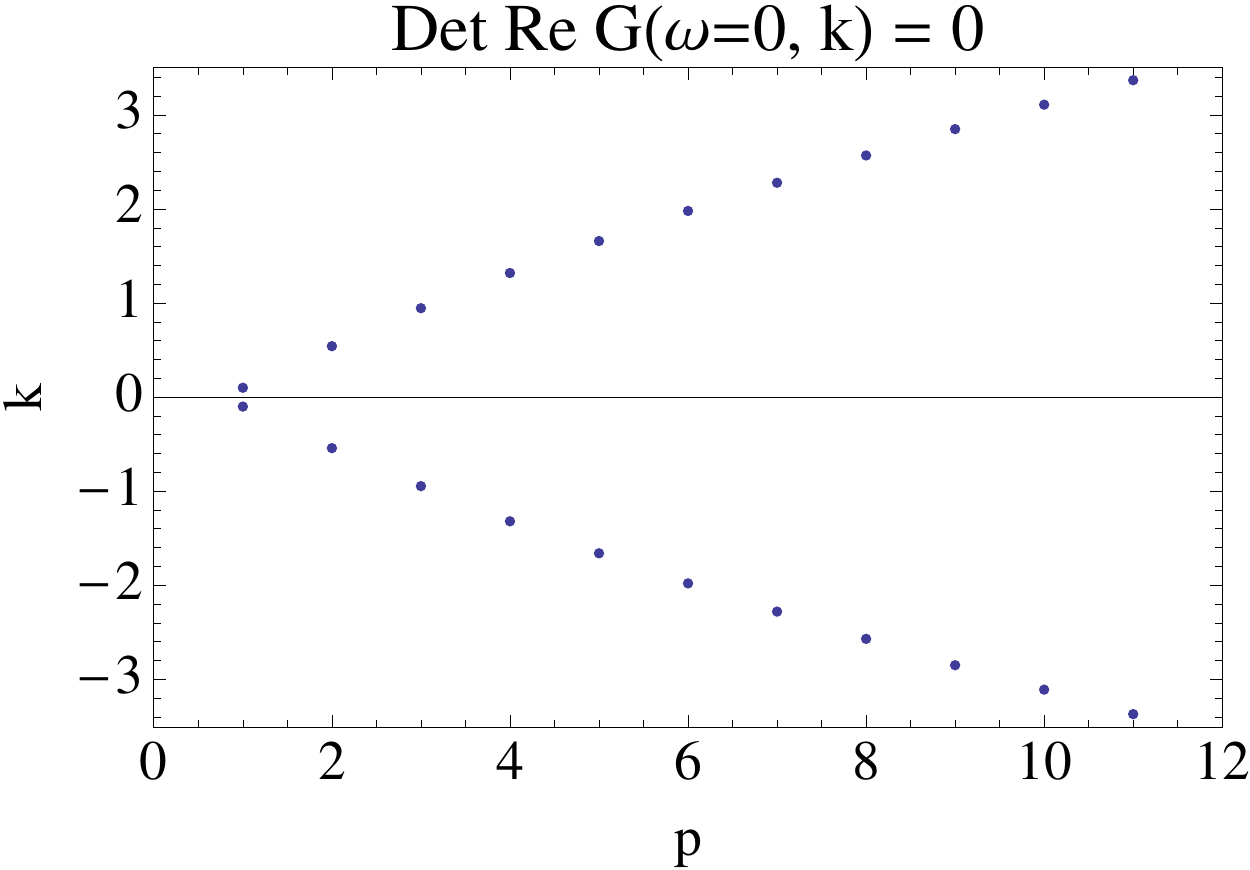}
	\end{subfigure}
	\captionsetup{justification=raggedright,singlelinecheck=false}
		\caption{\footnotesize Left: Determinant of the real part of $G$ at $\omega =0$. When $p=8$ (blue) we see a pair of zeros at $k \approx 2.57$. When $p=-8$ (red) we see poles at the same momenta. Right: Location of the zeros shown in the left panel as a function of $p$.}
	\label{zeros}
\end{figure*}

The divergence of the denominator is consistent with zeros of the retarded Green function  \cite{dzy,pzeros,zeros2,alsup}.  To lay  this plain, we plot ${\rm Det ReG}(\omega=0,k;p=\pm 8)$.   Indeed, 
${\rm Det Re G}(\omega=0,k;p=8)$ exhibits zeros precisely at the momenta at which ${\rm Det Re G}(\omega=0,k;p=-8)$ exhibits poles  as found earlier  \cite{alsup} for the RN-AdS$_4$background.  A plot of the zero surface as a function of $p$ is shown in the second panel in Fig. (\ref{zeros}).  Hence, zeros are the root cause of the gap. To understand the duality between zeros and poles,   consider the simpler case of this duality noted earlier\cite{faulkner}:  $G(\omega,k;-m)= -1/G(\omega,-k;m)$.  Because the solutions to Eq. (\ref{eq:asymp_soln}) scale as $r^{\pm mL}$, changing the sign of $m$ interchanges the role of the source and the response components of the retarded Green function.  Hence, if $G_-$ has poles then $G_+$ will have zeros and ${\rm Det G_R}(\omega,k,m)=1$.    The Pauli term also couples to the scaling dimension and as a consequence, the source and response at the boundary are sensitive to the sign of $p$.  To make this concrete, note that the solutions to the flow equations, Eq. (\ref{eq:flow}),  under the transformation $m\rightarrow -m$, $k\rightarrow -k$, and $p\rightarrow -p$ become $-1/\xi_{\pm}$ and as a consequence the duality between zeros and poles in the presence of the Pauli term is identical to the $p=0$ case and is just due to the switching between the role of the source and the response at the boundary.   Table \ref{table} contains the general form of the zeros-pole duality for arbitrary $(k,m,p)$.
The duality noted previously\cite{alsup} is the $m=0$ case.  
Since the propagator for $p$ sufficiently negative has only poles\cite{alsup}, zeros are inescapable for large positive values of $p$ and ${\rm Det G_R}(\omega=0,k;m,p)$ can vanish, a state of affairs which is not possible for the $p=0$ case.    As a gap resulting from zeros is a robust feature,  the Pauli term offers a generic way of modeling Mott physics in holographic settings independent of the Reissner-Nordstr\"om geometry.  There is a curious feature of the Pauli term, however. Because of the zeros-poles duality, the charge density determined even in the Mott insulating phase is determined by the sign changes of the Green function and hence satisfies the Luttinger theorem.  This appears to be an accident because in general as long as $SU(N)$ symmetry is intact, the Luttinger sum rule does not apply  \cite{nlutt} to systems  in which the self energy diverges as in Mott insulators.  
\begin{table}
\begin{tabular}{|c|c|c|}		
                     \hline
			{\rm Parameter choices} & $G_{\pm} (\omega, k; m,p)$ & ${\rm Det G_R} (\omega, k; m,p) $\\
		\hline
			$\begin{array}[c]{@{}c@{}}k \leftrightarrow -k\end{array}$& $G_{\mp} (\omega, -k; m,p)$ & \text{---} \\
		\hline
			$\begin{array}[c]{@{}c@{}}m = 0\\ p = 0\end{array}$& $\frac{-1}{G_{\pm} (\omega, -k)}$ & 1 \\
		\hline
			$\begin{array}[c]{@{}c@{}}m \neq 0\\p \neq 0\end{array}$& $\frac{-1}{G_{\pm}(\omega, -k; -m, -p) }$&$\frac{1}{{\rm Det G_R} (\omega, k; -m, -p)}$\\
		\hline
		\end{tabular}
			\captionsetup{justification=raggedright,singlelinecheck=false}
	\caption{\footnotesize Table showing the duality relating the components and the determinant of the Green function under $k\rightarrow -k$, $m\rightarrow -m$, $p\rightarrow -p$.  For the case of $m=0$, the determinant of $G^R$ is restricted to be unity, reflecting the fact that the poles and zeros exactly cancel.  For small values of $|p|$, $G_{\pm}$ has both poles and zeros.  However, for $p<-p_c$, only poles exist implying that for $p>p_c$ zeros obtain regardless of the value of the mass $m$.  $p_c$ is background dependent.  For RN-AdS, $p_c\approx 4.0$, whereas for the and SS-AdS, $p_c\approx 6.8$ for $Q=\sqrt{3}$. }
	\label{table}
	\end{table}

\section{Final Remarks}

Our work here establishes quite generally that holographic constructions with the Pauli term result in zeros of the single-particle Green function and hence can be used to study non-perturbative Mott physics.  In fact, coupled with the supergravity string result\cite{gauntlett,gubser2} that spectral functions do not  generally display Fermi surfaces, the gapping tendency of the Pauli term could provide a simple way of modeling this 
physics.  Hence, it could offer a new way of modeling non-perturbative fermionic matter at finite density.  The tethering of the gap to the Pauli term and zeros implies that the Pauli term increases the scaling dimension sufficiently of the boundary operators such that the $\rm Det G_R(0)=0$ rather than $\rm Det G_R(0)=1$, as in the Fermi liquid case.  This implies that the generic form of the propagator is consistent with the unparticle kind with a large anomalous scaling dimension as we have proposed previously  \cite{unparticle}.   Precisely how such large scaling dimensions arise in more microscopic theories is a topic for future study.   The zeros-poles duality found here in  a holographic setting ensures that a divergent self energy can be recast as a dual Fermi surface problem.  We are unaware of any condensed matter model which exhibits this duality.  Perhaps if one could be constructed, this would greatly expand our understanding of the Mott problem.  However, in a pure condensed matter setting, the very notion of standard and alternate quantization for the retarded Green function does not exist and hence such a duality between zeros and poles seems unlikely.

%%%%%%%%%%%%%%%%%%%%%%

\textbf{Acknowlegements:} We thank J. Alsup for an email exchange on their unpublished results, Tom Faulkner and Mohammad Edalati for their characteristically level-headed remarks and the NSF DMR-1104909 for partial funding of this project. This work grew out of work funded earlier by the
Center for Emergent Superconductivity, a DOE Energy Frontier Research Center, Grant No.~DE-AC0298CH1088.

\section*{Appendix: Boundary Conditions}

The in-falling boundary conditions on $\psi_{\pm}$ are most easily determined from the equations of motion for their components. Here we work only in the $m=0$ case studied in this paper. Substituting $\psi_{\pm} = (\beta_{\pm}, \alpha_{\pm})^T$ into the Dirac equation \eqref{eq:eom} (in its dimensionless variable form) and decoupling the two resultant equations gives
	\begin{align} \label{eq:alpha}
		- \frac{u^2 \sqrt{f(u)}}{v_{-}(u) \mp k} \partial_u \left( \frac{u^2 \sqrt{f(u)}}{v_{+}(u) \pm k} \partial_u \right) \alpha_{\pm} &= \alpha_{\pm}, \\ \label{eq:beta}
		- \frac{u^2 \sqrt{f(u)}}{v_{+}(u) \pm k} \partial_u \left( \frac{u^2 \sqrt{f(u)}}{v_{-}(u) \mp k} \partial_u \right) \beta_{\pm} &= \beta_{\pm},
	\end{align}
where $v_{\pm} (u)$ are defined as in \eqref{eq:v}. When $\omega \neq 0$ these equations expand to simple scaling forms around the horizon, and we find that the general solutions are
	\begin{align}
		\alpha_- (u) \approx A_1 e^{i(\omega/d) \ln (u-1)} + A_2 e^{-i(\omega/d) \ln (u-1)},
	\end{align}
and similarly for $\beta_- (u)$, where $A_1$ and $A_2$ are undetermined phases. We retain the in-falling solutions and use the flow equation \eqref{eq:flow} to solve for the phase of $\beta_{\pm}$ in terms of the phase of $\alpha_{\pm}$, finding $B = i A$ or
	\begin{align}
		\xi_{\pm} \big|_{u=1} = i \qquad \omega \neq 0.
	\end{align}

When $\omega = 0$ the notion of in-falling is lost, so we use regularity to constrain near-horizon solutions for the components. In this case the near-horizon expansion of \eqref{eq:alpha} is
	\begin{align} \label{eq:anh}
		\left( C \sqrt{x} - D x \right) \partial_x \left[ \left( C \sqrt{x} + D x \right) \alpha'_- (u) \right] = \alpha_- (u),
	\end{align}
where $x \equiv (u - 1)$ and
	\begin{align}
		C \equiv \frac{\sqrt{d}}{(k - pQ)}, \quad D \equiv \frac{(d-2)Q}{(k -pQ)^2}.
	\end{align}
The near-horizon equation \eqref{eq:anh} may be cast in hypergeometric form through successive coordinate transformations to $y \equiv D \sqrt{x}/C$ and $z \equiv (1-y)/2$. The general solutions for the components are
	\begin{align}
		\alpha_{-} (y) &= A_1 (y-1)  {_2F_1} \left[1- \tfrac{2i}{|D|}, 1+ \tfrac{2i}{|D|}; 2; \tfrac{1-y}{2} \right] \nonumber \\
			&\quad+ A_2 G^{2,0}_{2,2} \left[ \frac{1-y}{2} \  \bigg|
				\begin{array}{cc}
					1 - \tfrac{2i}{|D|} & 1+ \tfrac{2i}{|D|} \\
					0 & 1
				\end{array}
		 \right], \label{eq:abc}
	\end{align}
	\begin{align}
		\beta_{-} (y) &= B_1 \   {_2F_1} \left[- \tfrac{2i}{|D|}, \tfrac{2i}{|D|}; 1; \tfrac{1-y}{2} \right] \nonumber \\
			&\quad + B_2 G^{2,0}_{2,2} \left[ \frac{1-y}{2} \  \bigg|
				\begin{array}{cc}
					1 - \tfrac{2i}{|D|} & 1+ \tfrac{2i}{|D|} \\
					0 & 0
				\end{array}
		 \right] \label{eq:bbc},
	\end{align}
where $_2 F_1$ is a hypergeometric function, $G_{2,2}^{2,0}$ is a Meijer G-function, and $A_1, A_2, B_1$, and $B_2$ are undetermined phases. The G-functions are divergent at the horizon, so we retain only the hypergeometric functions. Thus $\xi_-$ is given by a ratio of these functions and \eqref{eq:flow} defines a quadratic equation for $B_1/A_1$. An appropriate choice of the root gives a numerically computable boundary condition on $\xi_-$ which, when used to integrate \eqref{eq:flow}, produces boundary Green functions consistent with those obtained from \eqref{eq:alpha} and \eqref{eq:beta}  subject to \eqref{eq:abc} and \eqref{eq:bbc}. 

Since we are concerned with the presence of spectral weight at $\omega =0$ -- which only occurs when the boundary condition on $\xi_-$ has an imaginary part -- we do not provide the exact form of the previously discussed roots. In the results reported above we found all zero frequency boundary conditions on $\xi_-$ to be real, meaning there is no contribution to the spectral weight. In the $\omega = 0$ plots of ${\rm Det} \re G$ we used the larger of the two roots as boundary data, as we find that it generally leads to more tractable numerics.
%\bibliography{zpduality}

\begin{thebibliography}{10}%
\makeatletter
\providecommand \@ifxundefined [1]{%
 \ifx #1\undefined \expandafter \@firstoftwo
 \else \expandafter \@secondoftwo
\fi
}%
\providecommand \@ifnum [1]{%
 \ifnum #1\expandafter \@firstoftwo
 \else \expandafter \@secondoftwo
\fi
}%
\providecommand \enquote [1]{``#1''}%
\providecommand \bibnamefont  [1]{#1}%
\providecommand \bibfnamefont [1]{#1}%
\providecommand \citenamefont [1]{#1}%
\providecommand\href[0]{\@sanitize\@href}%
\providecommand\@href[1]{\endgroup\@@startlink{#1}\endgroup\@@href}%
\providecommand\@@href[1]{#1\@@endlink}%
\providecommand \@sanitize [0]{\begingroup\catcode`\&12\catcode`\#12\relax}%
\@ifxundefined \pdfoutput {\@firstoftwo}{%
 \@ifnum{\z@=\pdfoutput}{\@firstoftwo}{\@secondoftwo}%
}{%
 \providecommand\@@startlink[1]{\leavevmode}%
 \providecommand\@@endlink[0]{}%
}{%
 \providecommand\@@startlink[1]{%
  \leavevmode
  \pdfstartlink
   attr{/Border[0 0 1 ]/H/I/C[0 1 1]}%
   user{/Subtype/Link/A<</Type/Action/S/URI/URI(#1)>>}%
  \relax
 }%
 \providecommand\@@endlink[0]{\pdfendlink}%
}%
\providecommand \url  [0]{\begingroup\@sanitize \@url }%
\providecommand \@url [1]{\endgroup\@href {#1}{\urlprefix}}%
\providecommand \urlprefix [0]{URL }%
\providecommand \Eprint[0]{\href }%
\@ifxundefined \urlstyle {%
  \providecommand \doi [1]{doi:\discretionary{}{}{}#1}%
}{%
  \providecommand \doi [0]{doi:\discretionary{}{}{}\begingroup
  \urlstyle{rm}\Url }%
}%
\providecommand \doibase [0]{http://dx.doi.org/}%
\providecommand \Doi[1]{\href{\doibase#1}}%
\providecommand \bibAnnote [3]{%
  \BibitemShut{#1}%
  \begin{quotation}\noindent
    \textsc{Key:}\ #2\\\textsc{Annotation:}\ #3%
  \end{quotation}%
}%
\providecommand \bibAnnoteFile [2]{%
  \IfFileExists{#2}{\bibAnnote {#1} {#2} {\input{#2}}}{}%
}%
\providecommand \typeout [0]{\immediate \write \m@ne }%
\providecommand \selectlanguage [0]{\@gobble}%
\providecommand \bibinfo [0]{\@secondoftwo}%
\providecommand \bibfield [0]{\@secondoftwo}%
\providecommand \translation [1]{[#1]}%
\providecommand \BibitemOpen[0]{}%
\providecommand \bibitemStop [0]{}%
\providecommand \bibitemNoStop [0]{.\EOS\space}%
\providecommand \EOS [0]{\spacefactor3000\relax}%
\providecommand \BibitemShut [1]{\csname bibitem#1\endcsname}%
%</preamble>
\bibitem{alsup}%
  \BibitemOpen
  \bibfield{author}{%
  \bibinfo {author} {\bibfnamefont{G.~S. K.~Y.}\ \bibnamefont{J.~Alsup},
  \bibfnamefont{E.~Papantonopoulos}},\ }%
  \bibinfo {journal} {http://arxiv.org/abs/1404.4010}%
  \bibAnnoteFile{NoStop}{alsup}%
\bibitem{gauntlett}%
  \BibitemOpen
\bibfield{journal}{%
    }%
  \bibfield{author}{%
  \bibinfo {author} {\bibfnamefont{J.~P.}\ \bibnamefont{Gauntlett}}, \bibinfo
  {author} {\bibfnamefont{J.}~\bibnamefont{Sonner}},\ and\ \bibinfo {author}
  {\bibfnamefont{D.}~\bibnamefont{Waldram}},\ }%
  \bibfield{journal}{%
  \Doi{10.1103/PhysRevLett.107.241601}{\bibinfo {journal} {Phys. Rev. Lett.}}\
  }%
  \textbf{\bibinfo {volume} {107}},\ \bibinfo {pages} {241601} (\bibinfo
  {month} {Dec}\ \bibinfo {year} {2011}),\
  \url{http://link.aps.org/doi/10.1103/PhysRevLett.107.241601}%
  \bibAnnoteFile{NoStop}{gauntlett}%
\bibitem{gubser2}%
  \BibitemOpen
  \bibfield{author}{%
  \bibinfo {author} {\bibfnamefont{O.}~\bibnamefont{DeWolfe}}, \bibinfo
  {author} {\bibfnamefont{S.~S.}\ \bibnamefont{Gubser}},\ and\ \bibinfo
  {author} {\bibfnamefont{C.}~\bibnamefont{Rosen}},\ }%
  \bibfield{journal}{%
  \Doi{10.1103/PhysRevLett.108.251601}{\bibinfo {journal} {Phys.Rev.Lett.}}\ }%
  \textbf{\bibinfo {volume} {108}},\ \bibinfo {pages} {251601} (\bibinfo {year}
  {2012}),\ \Eprint{http://arxiv.org/abs/1112.3036}{arXiv:1112.3036 [hep-th]}%
  \bibAnnoteFile{NoStop}{gubser2}%
%%CITATION = ARXIV:1112.3036;%%
\bibitem{thooft}%
  \BibitemOpen
  \bibfield{author}{%
  \bibinfo {author} {\bibnamefont{G.}}\ and\ \bibinfo {author}
  {\bibnamefont{'t~Hooft}},\ }%
  \bibfield{journal}{%
  \Doi{10.1016/0550-3213(74)90088-1}{\bibinfo {journal} {Nuclear Physics B}}\
  }%
  \textbf{\bibinfo {volume} {75}},\ \bibinfo {pages} {461 } (\bibinfo {year}
  {1974}),\ ISSN \bibinfo {issn} {0550-3213},\
  \url{http://www.sciencedirect.com/science/article/pii/0550321374900881}%
  \bibAnnoteFile{NoStop}{thooft}%
\bibitem{dzy}%
  \BibitemOpen
  \bibfield{author}{%
  \bibinfo {author} {\bibfnamefont{I.}~\bibnamefont{Dzyaloshinskii}},\ }%
  \bibfield{journal}{%
  \Doi{10.1103/PhysRevB.68.085113}{\bibinfo {journal} {Phys. Rev. B}}\ }%
  \textbf{\bibinfo {volume} {68}},\ \bibinfo {pages} {085113} (\bibinfo {month}
  {Aug}\ \bibinfo {year} {2003})%
  \bibAnnoteFile{NoStop}{dzy}%
\bibitem{dinfty}%
  \BibitemOpen
  \bibfield{author}{%
  \bibinfo {author} {\bibfnamefont{A.}~\bibnamefont{Georges}}, \bibinfo
  {author} {\bibfnamefont{G.}~\bibnamefont{Kotliar}}, \bibinfo {author}
  {\bibfnamefont{W.}~\bibnamefont{Krauth}},\ and\ \bibinfo {author}
  {\bibfnamefont{M.~J.}\ \bibnamefont{Rozenberg}},\ }%
  \bibfield{journal}{%
  \Doi{10.1103/RevModPhys.68.13}{\bibinfo {journal} {Rev. Mod. Phys.}}\ }%
  \textbf{\bibinfo {volume} {68}},\ \bibinfo {pages} {13} (\bibinfo {month}
  {Jan}\ \bibinfo {year} {1996})%
  \bibAnnoteFile{NoStop}{dinfty}%
\bibitem{kotstan}%
  \BibitemOpen
  \bibfield{author}{%
  \bibinfo {author} {\bibfnamefont{T.~D.}\ \bibnamefont{Stanescu}}\ and\
  \bibinfo {author} {\bibfnamefont{G.}~\bibnamefont{Kotliar}},\ }%
  \bibfield{journal}{%
  \Doi{10.1103/PhysRevB.74.125110}{\bibinfo {journal} {Phys. Rev. B}}\ }%
  \textbf{\bibinfo {volume} {74}},\ \bibinfo {pages} {125110} (\bibinfo {month}
  {Sep}\ \bibinfo {year} {2006}),\
  \url{http://link.aps.org/doi/10.1103/PhysRevB.74.125110}%
  \bibAnnoteFile{NoStop}{kotstan}%
\bibitem{imada}%
  \BibitemOpen
  \bibfield{author}{%
  \bibinfo {author} {\bibfnamefont{H.}~\bibnamefont{Park}}, \bibinfo {author}
  {\bibfnamefont{K.}~\bibnamefont{Haule}},\ and\ \bibinfo {author}
  {\bibfnamefont{G.}~\bibnamefont{Kotliar}},\ }%
  \bibfield{journal}{%
  \bibinfo {journal} {Physical Review Letters}\ }%
  \textbf{\bibinfo {volume} {101}},\ \bibinfo {pages} {186403} (\bibinfo {year}
  {2008}),\ \url{http://link.aps.org/abstract/PRL/v101/e186403}%
  \bibAnnoteFile{NoStop}{imada}%
\bibitem{maldacena}%
  \BibitemOpen
  \bibfield{author}{%
  \bibinfo {author} {\bibfnamefont{J.}~\bibnamefont{Maldacena}},\ }%
  \bibfield{journal}{%
  \bibinfo {journal} {Adv. Theor. Math. Phys.}\ }%
  \textbf{\bibinfo {volume} {2}},\ \bibinfo {pages} {231} (\bibinfo {year}
  {1998})%
  \bibAnnoteFile{NoStop}{maldacena}%
\bibitem{witten}%
  \BibitemOpen
  \bibfield{author}{%
  \bibinfo {author} {\bibfnamefont{E.}~\bibnamefont{Witten}},\ }%
  \bibfield{journal}{%
  \bibinfo {journal} {Theor. Math. Phys.}\ }%
  \textbf{\bibinfo {volume} {2}},\ \bibinfo {pages} {505} (\bibinfo {year}
  {1998})%
  \bibAnnoteFile{NoStop}{witten}%
\bibitem{gubser}%
  \BibitemOpen
  \bibfield{author}{%
  \bibinfo {author} {\bibfnamefont{I.~R.~K.}\ \bibnamefont{S.~S.~Gubser}}\ and\
  \bibinfo {author} {\bibfnamefont{A.~M.}\ \bibnamefont{Polyakov}},\ }%
  \bibfield{journal}{%
  \bibinfo {journal} {Phys. Lett. B}\ }%
  \textbf{\bibinfo {volume} {428}},\ \bibinfo {pages} {105} (\bibinfo {year}
  {1998})%
  \bibAnnoteFile{NoStop}{gubser}%
\bibitem{edalati1}%
  \BibitemOpen
  \bibfield{author}{%
  \bibinfo {author} {\bibfnamefont{M.}~\bibnamefont{Edalati}}, \bibinfo
  {author} {\bibfnamefont{R.~G.}\ \bibnamefont{Leigh}},\ and\ \bibinfo {author}
  {\bibfnamefont{P.~W.}\ \bibnamefont{Phillips}},\ }%
  \bibfield{journal}{%
  \Doi{10.1103/PhysRevLett.106.091602}{\bibinfo {journal} {Phys. Rev. Lett.}}\
  }%
  \textbf{\bibinfo {volume} {106}},\ \bibinfo {pages} {091602} (\bibinfo
  {month} {Mar}\ \bibinfo {year} {2011}),\
  \url{http://link.aps.org/doi/10.1103/PhysRevLett.106.091602}%
  \bibAnnoteFile{NoStop}{edalati1}%
\bibitem{edalati2}%
  \BibitemOpen
  \bibfield{author}{%
  \bibinfo {author} {\bibfnamefont{M.}~\bibnamefont{Edalati}}, \bibinfo
  {author} {\bibfnamefont{R.~G.}\ \bibnamefont{Leigh}}, \bibinfo {author}
  {\bibfnamefont{K.~W.}\ \bibnamefont{Lo}},\ and\ \bibinfo {author}
  {\bibfnamefont{P.~W.}\ \bibnamefont{Phillips}},\ }%
  \bibfield{journal}{%
  \Doi{10.1103/PhysRevD.83.046012}{\bibinfo {journal} {Phys. Rev. D}}\ }%
  \textbf{\bibinfo {volume} {83}},\ \bibinfo {pages} {046012} (\bibinfo {month}
  {Feb}\ \bibinfo {year} {2011}),\
  \url{http://link.aps.org/doi/10.1103/PhysRevD.83.046012}%
  \bibAnnoteFile{NoStop}{edalati2}%
\bibitem{dipole1}%
  \BibitemOpen
  \bibfield{author}{%
  \bibinfo {author} {\bibfnamefont{J.-P.}\ \bibnamefont{Wu}},\ }%
  \bibfield{journal}{%
  \Doi{http://dx.doi.org/10.1016/j.physletb.2013.11.040}{\bibinfo {journal}
  {Physics Letters B}}\ }%
  \textbf{\bibinfo {volume} {728}},\ \bibinfo {pages} {450} (\bibinfo {month}
  {1}\ \bibinfo {year} {2014}),\
  \url{http://www.sciencedirect.com/science/article/pii/S037026931300943X}%
  \bibAnnoteFile{NoStop}{dipole1}%
\bibitem{dipole2}%
  \BibitemOpen
  \bibfield{author}{%
  \bibinfo {author} {\bibfnamefont{W.-Y.}\ \bibnamefont{{Wen}}}\ and\ \bibinfo
  {author} {\bibfnamefont{S.-Y.}\ \bibnamefont{{Wu}}},\ }%
  \bibfield{journal}{%
  \Doi{10.1142/S2010194513009525}{\bibinfo {journal} {International Journal of
  Modern Physics Conference Series}}\ }%
  \textbf{\bibinfo {volume} {21}},\ \bibinfo {pages} {143} (\bibinfo {month}
  {Jan.}\ \bibinfo {year} {2013})%
  \bibAnnoteFile{NoStop}{dipole2}%
\bibitem{dipole3}%
  \BibitemOpen
  \bibfield{author}{%
  \bibinfo {author} {\bibfnamefont{X.-M.}\ \bibnamefont{{Kuang}}}, \bibinfo
  {author} {\bibfnamefont{B.}~\bibnamefont{{Wang}}},\ and\ \bibinfo {author}
  {\bibfnamefont{J.-P.}\ \bibnamefont{{Wu}}},\ }%
  \bibfield{journal}{%
  \Doi{10.1007/JHEP07(2012)125}{\bibinfo {journal} {Journal of High Energy
  Physics}}\ }%
  \textbf{\bibinfo {volume} {7}},\ \bibinfo {pages} {125} (\bibinfo {month}
  {Jul.}\ \bibinfo {year} {2012}),\
  \Eprint{http://arxiv.org/abs/1205.6674}{arXiv:1205.6674 [hep-th]}%
  \bibAnnoteFile{NoStop}{dipole3}%
\bibitem{dipole4}%
  \BibitemOpen
  \bibfield{author}{%
  \bibinfo {author} {\bibfnamefont{W.-J.}\ \bibnamefont{{Li}}}\ and\ \bibinfo
  {author} {\bibfnamefont{H.}~\bibnamefont{{Zhang}}},\ }%
  \bibfield{journal}{%
  \Doi{10.1007/JHEP11(2011)018}{\bibinfo {journal} {Journal of High Energy
  Physics}}\ }%
  \textbf{\bibinfo {volume} {11}},\ \bibinfo {pages} {18} (\bibinfo {month}
  {Nov.}\ \bibinfo {year} {2011}),\
  \Eprint{http://arxiv.org/abs/1110.4559}{arXiv:1110.4559 [hep-th]}%
  \bibAnnoteFile{NoStop}{dipole4}%
\bibitem{dipole5}%
  \BibitemOpen
  \bibfield{author}{%
  \bibinfo {author} {\bibfnamefont{J.-P.}\ \bibnamefont{{Wu}}}\ and\ \bibinfo
  {author} {\bibfnamefont{H.-B.}\ \bibnamefont{{Zeng}}},\ }%
  \bibfield{journal}{%
  \Doi{10.1007/JHEP04(2012)068}{\bibinfo {journal} {Journal of High Energy
  Physics}}\ }%
  \textbf{\bibinfo {volume} {4}},\ \bibinfo {pages} {68} (\bibinfo {month}
  {Apr.}\ \bibinfo {year} {2012}),\
  \Eprint{http://arxiv.org/abs/1201.2485}{arXiv:1201.2485 [hep-th]}%
  \bibAnnoteFile{NoStop}{dipole5}%
\bibitem{dipole6}%
  \BibitemOpen
  \bibfield{author}{%
  \bibinfo {author} {\bibfnamefont{J.-P.}\ \bibnamefont{{Wu}}},\ }%
  \bibfield{journal}{%
  \Doi{10.1007/JHEP04(2013)073}{\bibinfo {journal} {Journal of High Energy
  Physics}}\ }%
  \textbf{\bibinfo {volume} {4}},\ \bibinfo {pages} {73} (\bibinfo {month}
  {Apr.}\ \bibinfo {year} {2013})%
  \bibAnnoteFile{NoStop}{dipole6}%
\bibitem{faulkner}%
  \BibitemOpen
  \bibfield{author}{%
  \bibinfo {author} {\bibfnamefont{T.}~\bibnamefont{{Faulkner}}}, \bibinfo
  {author} {\bibfnamefont{H.}~\bibnamefont{{Liu}}}, \bibinfo {author}
  {\bibfnamefont{J.}~\bibnamefont{{McGreevy}}},\ and\ \bibinfo {author}
  {\bibfnamefont{D.}~\bibnamefont{{Vegh}}},\ }%
  \bibfield{journal}{%
  \Doi{10.1103/PhysRevD.83.125002}{\bibinfo {journal} {\prd}}\ }%
  \textbf{\bibinfo {volume} {83}},\ \bibinfo {eid} {125002} (\bibinfo {month}
  {Jun.}\ \bibinfo {year} {2011}),\
  \Eprint{http://arxiv.org/abs/0907.2694}{arXiv:0907.2694 [hep-th]}%
  \bibAnnoteFile{NoStop}{faulkner}%
\bibitem{pzeros}%
  \BibitemOpen
  \bibfield{author}{%
  \bibinfo {author} {\bibfnamefont{T.~D.}\ \bibnamefont{Stanescu}}, \bibinfo
  {author} {\bibfnamefont{P.}~\bibnamefont{Phillips}},\ and\ \bibinfo {author}
  {\bibfnamefont{T.-P.}\ \bibnamefont{Choy}},\ }%
  \bibfield{journal}{%
  \Doi{10.1103/PhysRevB.75.104503}{\bibinfo {journal} {Physical Review B
  (Condensed Matter and Materials Physics)}}\ }%
  \textbf{\bibinfo {volume} {75}},\ \bibinfo {eid} {104503} (\bibinfo {year}
  {2007}),\ \url{http://link.aps.org/abstract/PRB/v75/e104503}%
  \bibAnnoteFile{NoStop}{pzeros}%
\bibitem{zeros2}%
  \BibitemOpen
  \bibfield{author}{%
  \bibinfo {author} {\bibfnamefont{F.~H.~L.}\ \bibnamefont{Essler}}\ and\
  \bibinfo {author} {\bibfnamefont{A.~M.}\ \bibnamefont{Tsvelik}},\ }%
  \bibfield{journal}{%
  \Doi{10.1103/PhysRevB.65.115117}{\bibinfo {journal} {Phys. Rev. B}}\ }%
  \textbf{\bibinfo {volume} {65}},\ \bibinfo {pages} {115117} (\bibinfo {month}
  {Mar}\ \bibinfo {year} {2002})%
  \bibAnnoteFile{NoStop}{zeros2}%
\bibitem{nlutt}%
  \BibitemOpen
  \bibfield{author}{%
  \bibinfo {author} {\bibfnamefont{K.~B.}\ \bibnamefont{{Dave}}}, \bibinfo
  {author} {\bibfnamefont{P.~W.}\ \bibnamefont{{Phillips}}},\ and\ \bibinfo
  {author} {\bibfnamefont{C.~L.}\ \bibnamefont{{Kane}}},\ }%
  \bibfield{journal}{%
  \Doi{10.1103/PhysRevLett.110.090403}{\bibinfo {journal} {Physical Review
  Letters}}\ }%
  \textbf{\bibinfo {volume} {110}},\ \bibinfo {eid} {090403} (\bibinfo {month}
  {Mar.}\ \bibinfo {year} {2013}),\
  \Eprint{http://arxiv.org/abs/1207.4201}{arXiv:1207.4201 [cond-mat.str-el]}%
  \bibAnnoteFile{NoStop}{nlutt}%
\bibitem{unparticle}%
  \BibitemOpen
  \bibfield{author}{%
  \bibinfo {author} {\bibfnamefont{P.~W.}\ \bibnamefont{{Phillips}}}, \bibinfo
  {author} {\bibfnamefont{B.~W.}\ \bibnamefont{{Langley}}},\ and\ \bibinfo
  {author} {\bibfnamefont{J.~A.}\ \bibnamefont{{Hutasoit}}},\ }%
  \bibfield{journal}{%
  \Doi{10.1103/PhysRevB.88.115129}{\bibinfo {journal} {\prb}}\ }%
  \textbf{\bibinfo {volume} {88}},\ \bibinfo {eid} {115129} (\bibinfo {month}
  {Sep.}\ \bibinfo {year} {2013}),\
  \Eprint{http://arxiv.org/abs/1305.0006}{arXiv:1305.0006 [cond-mat.str-el]}%
  \bibAnnoteFile{NoStop}{unparticle}%
\end{thebibliography}
%Merlin.mbs v4.21 2009-07-09.
%
\end{document}